\documentclass[conference]{IEEEtran}

\usepackage[dvips]{graphics}
\usepackage{graphicx}
\usepackage{times}
\usepackage{amsmath}
\usepackage{amssymb}
\usepackage{fancybox}
\usepackage[usenames,dvipsnames]{color}

\newtheorem{lemma}{\bf Lemma}
\newtheorem{theorem}{\bf Theorem}

\newcommand{\bee}{\begin{eqnarray}}
\newcommand{\eee}{\end{eqnarray}}
\newcommand{\be}{\begin{equation}}
\newcommand{\ee}{\end{equation}}
\newcommand{\al}[1]{\begin{align} #1 \end{align}}

\newcommand{\equ}[1]{\begin{equation} #1 \end{equation}}
\newcommand{\mb}{\mathbf}

\newcommand{\bs}{\boldsymbol}

\newcommand{\nnb}{\nonumber}
\newcommand{\qa}{{\bf a}}

\newcommand{\qg}{{\bf g}}
\newcommand{\qh}{{\bf h}}
\newcommand{\qn}{{\bf n}}

\newcommand{\qA}{{\bf A}}

\newcommand{\qG}{{\bf G}}
\newcommand{\qH}{{\bf H}}
\newcommand{\qI}{{\bf I}}

\begin{document}

\title{Physical Layer Security with Uncoordinated Helpers Implementing Cooperative Jamming }
\author{Shuangyu Luo, Jiangyuan Li, and Athina Petropulu\\Rutgers-The State University of New Jersey, Piscataway, NJ 08854}
\maketitle

\begin{abstract}

A wireless communication network is considered, consisting of a source (Alice), a destination (Bob) and an eavesdropper (Eve), each equipped with a single antenna. The communication is assisted by multiple helpers, each equipped with two antennas, which implement cooperative jamming, i.e., transmitting noise to confound Eve. The optimal structure of the jamming noise that maximizes the secrecy rate is derived. A nulling noise scenario is also considered, in which each helper transmits noise that nulls out at Bob. Each helper only requires knowledge of its own link to Bob to determine the noise locally. For the optimally structured noise, global information of all the links is required. Although analysis shows that under the two-antenna per helper scenario the nulling solution is sub-optimal in terms of the achievable secrecy rate, simulations show that the performance difference is rather small, with the inexpensive and easy to implement nulling scheme performing near optimal.
\end{abstract}


\section{Introduction}
\label{sec:Intro}


In wireless communication networks, security is an important issue since signals are broadcasted in the air. Physical layer security has drawn considerable attention since Wyner \cite{Wyner75} proposed the wiretap channel in 1975. \cite{poor} provided a theoretic security analysis using information theory. Consider a wiretap channel and a typical scenario in which Alice, the source node, wants to have a private conversation with Bob. The problem is that Eve can also get the broadcasted signal. By deploying some cooperative relays between Alice and Bob could help Alice to keep the signal secret. There are mainly three relaying methods: Decode-and-Forward, Amplify-and-Forward  and Cooperative Jamming (CJ).

Cooperative Jamming is a method that relays transmit noise to degrade the channels of eavesdropper.
In CJ, while the purpose of transmitting noise is to confound the eavesdropper, the noise unavoidably affects the legitimate receiver. In \cite{Dong2}, a weighted common noise is transmitted by each relay in a way to ensure nulling at the destination, and suboptimal but closed form  weights and relay power allocation were provided to maximize the secrecy rate subject to power constraints. As an improvement to \cite{Dong2}, \cite{Jiangyuan1} proposed optimal weight design and power allocation that do not rely on nulling at the destination. The aforementioned methods require global channel information at the relays, including that of the eavesdroppers.
On the other hand, \cite{Swindlehurst}, \cite{Goel1} proposed transmitting structured noise rather than weighted random noise. In \cite{Swindlehurst}, all nodes have multiple antennas, and the jamming could be coordinated (requiring public information), or uncoordinated (UCJ) (no public information).
In UCJ, the helpers minimize the interference to Bob autonomously by sending the jamming signals along the right singular vector that correspond to the smallest singular value of their channel to Bob. Again, no eavesdropper channel information is needed.

In this paper, we consider a scenario in which Alice, Bob and Eve are all equipped with a single antenna, while each relay is equipped with two antennas.
We study two artificial noise schemes, the local nulling artificial noise and the general structured artificial noise.
In local nulling, each helper completely cancels its interference at Bob, using only  local information of its link to Bob.
The condition for achieving a local nulling is that the artificial noise at each helper
can be designed to lie in the null space of its channel to Bob, which requires that the
 relay has at least two antennas. Taking cost, size limitation into consideration, we proceed with two antennas and but the result can be directly extended for
any number of antennas at relays. The optimal  structure for the jamming noise  that maximizes the secrecy rate is also derived.



We point out that local nulling is a special case of UCJ \cite{Swindlehurst}.
In \cite{Swindlehurst}, the UCJ for MIMO channel is studied,
each helper's interference to Bob is not necessarily cancelled locally;
this depends on the degrees of freedom of each helper's channel matrix.
For example, if each helper has two antennas and Bob has two antennas, then
each helper's channel matrix $\qH_i$ has a size of $2\times 2$.
For this case, there is not an artificial noise $\mb{q}_i \ne 0$ such that $\qH_i\mb{q}_i=0$.
However, local nulling is the case in which each helper's interference to Bob is always cancelled locally.

In the following, we  provide the system configuration in Section \ref{sec:System Model}. The nulling noise scenario is described in  Section \ref{subsec:Orthogonal Artificial Noise}, and optimally structured noise scenario in discussed in Section \ref{subsec:General Structured Artificial Noise}. In Section \ref{sec:Simulations}  we show that the nulling solution performs very close to the optimal solution in terms of secrecy rate. Finally, Section \ref{sec:Conclusions} provides some concluding remarks.

{\em Notation -} Throughout this paper, the following notations are adopted.
Upper case and lower case bold symbols denote matrices and vectors, respectively.
Superscripts $\ast$, $T$ and $\dagger$
denote respectively conjugate, transposition and conjugate transposition.
$\mathrm{Tr}({\mb A})$ denotes the trace of the matrix $\mb A$.
$\lambda_{\max}(\qA)$ denotes the largest eigenvalue of the matrix $\qA$.
${\mb A}\succeq 0$ denotes that the matrix $\qA$ is Hermitian positive semi-definite.
$|a|$ denotes absolute value of the complex number $a$.
$\|\qa\|=\sqrt{\qa^\dagger\qa}$ denotes Euclidean norm of the vector $\qa$.
$\qI_n$ denotes the identity matrix of order $n$ (the subscript is dropped when the dimension is obvious).
$\mathbb{E}\{\cdot\}$ denotes expectation operator.
$\mathrm{i}=\sqrt{-1}$.

\begin{figure}[hbtp]
\centering
\includegraphics[scale=0.8]{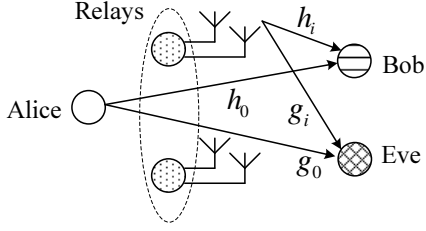}
\caption{System model.}
\label{systemmodel}
\end{figure}

\section{System Model and Problem Formulation}
\label{sec:System Model}

We consider a wireless network model consisting of a source node (Alice), $N$ trusted relays, a destination node (Bob) and a eavesdropper node (Eve),
as shown in Fig. \ref{systemmodel}.
Bob and Eve can only passively receive signals from source and relays, but not transmit signals.
 Alice, Bob, Eve, have antenna each and operates in half-duplex mode.
Each relay  has two antennas.
We denote the channels between $i$th relay and Bob, and  $i$th relay and Eve
as $\mathbf{h}_i$ and $\mathbf{g}_i$
(both $2\times 1$), respectively.
The channels Alice-Bob, and Alice-Eve are denoted by $h_0$ and $g_0$, respectively.
The source transmits a symbol $\sqrt{P_s}\, x$ with $\mathbb{E}\{\vert x \vert ^{2}\}=1$.
The individual power budget of relay $i$ is $P_i$.

A cooperative jamming scenario is considered, i.e., while Alice is transmitting,  the relays  transmit independent noise signals, which are independent of the source message.
In particular, relay $i$ transmits a $2\times 1$ noise vector $\mathbf{n}_i \sim \mathcal{CN}(0, P_i {\bs \Sigma}_i)$ with ${\bs \Sigma}_i\succeq 0$ and $\mathrm{Tr}({\bs \Sigma}_i)\le 1$, $i=1, \cdots, N$.
In the following, we address two different scenarios for the transmitted noise, i.e., the nulling noise and the optimally structured noise.
For the nulling noise case, the relay $i$ only requires  knowledge of its own link (i.e., $\qh_i$) and determines
the  noise locally.
For the optimally structured  noise, global information of all the links is required.

\subsection{Nulling Noise}
\label{subsec:Orthogonal Artificial Noise}

In this section, we  design  noise to lie in the null space of the channel, so that it will jam Eve but
cause no interference to Bob. A simple solution to this problem can be obtained.


In this scheme, only local information is required. Interestingly enough,  analysis and numerical simulation show that,
the nulling solution achieves a good approximation of the optimal solution in average sense.


For a noise vector $\mathbf{n}_i$, associated with node $i$, to cause  nulling at Bob, it must hold
\begin{equation}
\qh_i^T \mathbf{n}_i =0, \ i=1,\dots, N. \label{orthogonal}
\end{equation}
Let $\mathbf{h}_i^\perp \in \mathbb{C}^{2\times1}$ be an orthonormal basis for the null space of $\mathbf{h}_i^T$ with $\|\qh_i^\perp\|^2 = 1$.
We can express $\qn_i$ as
\begin{equation}
\qn_i=w_i \mathbf{h}_i^\perp v_i, \ i=1,\dots, N
\end{equation}
where $v_i$ is a zero mean Gaussian random random variable with unit covariance, $w_i$ is the weight to be determined.
The power of the noise is $\mathbb{E}\{\|\qn_i\|^2\} = |w_i|^2 \|\mathbf{h}_i^\perp\|^2\mathbb{E}\{|v_i|^2\} = |w_i|^2$.
In fact, this corresponds to $\mathbf{n}_i \sim \mathcal{CN}(0, |w_i|^2 \mathbf{h}_i^\perp (\mathbf{h}_i^\perp)^\dagger)$.
Using the fact $\qh_i^T\mathbf{h}_i^\perp=0$, $\forall i$, the received signal at Bob and Eve is $y_b$ and $y_e$ respectively:
\al{
y_b &= \sqrt{P_{s}}h_{0}x+ \sum_{i=1}^N  \qh_i^T(w_i \mathbf{h}_i^\perp v_i) +  n_b, \nnb\\
&= \sqrt{P_{s}}h_{0}x + n_b, \label{yb}\\
y_e &= \sqrt{P_{s}}g_{0}x+\sum_{i=1}^{N} \mathbf{g}_i^T (w_i \mathbf{h}_i^\perp v_i)  + n_e
}
where $n_b$ and $n_e$ are the AWGN received at Bob and Eve. For simplicity, assume $\mathbb{E}[|n_b|^2]=N_0$ and $\mathbb{E}[|n_e|^2]=N_0$.

The secrecy rate of the above system can be expressed as
\al{
&R_{1}(\{w_i\})= \log \left( 1+\frac{P_s|h_0|^2}{N_0}\right) \nnb\\
&\qquad -\log \left( 1+\frac{P_s|g_0|^2}{\sum_{i=1}^{N} |w_i|^2 \mathbf{g}_i^T \mathbf{h}_i^\perp(\mathbf{h}_i^\perp)^\dagger \mathbf{g}_i^\ast +N_0}\right). \label{Eq:RPerfect1}
}
Obviously, to maximize $R_1$, it should hold that  $|w_i|^2=P_i$, $\forall i$. Thus, the noise $\qn_i$ is given by
\begin{equation}
\qn_i= \sqrt{P_i}\ \mathbf{h}_i^\perp v_i, \ i=1,\dots, N
\end{equation}
and the secrecy rate equals
\al{
R_{1}=&\log \left( 1+\gamma_0 |h_0|^2\right)\nnb\\
&-\log \left( 1+\frac{\gamma_0|g_0|^2}{\sum_{i=1}^{N} \gamma_i \mathbf{g}_i^T \mathbf{h}_i^\perp(\mathbf{h}_i^\perp)^\dagger  \mathbf{g}_i^\ast +1}\right) \label{Eq:R1Pi}
}
where
\al{
\gamma_0 &= \frac{P_s}{N_0}, \label{gamma_0}\\
\gamma_i &= \frac{P_i}{N_0}, \ i=1, \cdots, N. \label{gamma_i}
}

\medskip

{\em Discussion}- The relays do not need information on the eavesdropper channel. Further, relay $i$ requires only the knowledge of its own link, i.e., $\qh_i$ and hence locally determines its weight.
As a result, such  system can be implemented in a distributed fashion, which greatly facilitates a real world implementation.

\subsection{General Structured Artificial Noise}
\label{subsec:General Structured Artificial Noise}

The above described nulling scheme is not optimal in terms of secrecy rate. One would not know
 how much secrecy rate is lost with the nulling scheme unless one compares that solution to the optimal one.
  In this section
  we derive  the optimal solution for the jamming noise. This solution will require global channel information available at each relay. Interestingly, it turns out that the added cost of the global channel information does not buy significant secrecy rate improvement, thus suggesting that the nulling solution might be preferable in a real world implementation.

 The received signal at Bob and Eve will be:
\begin{eqnarray}
y_b&=&\sqrt{P_{s}}\, h_{0}x+\sum_{i=1}^{N} \mathbf{h}_i^T \mathbf{n}_i+n_b,\\
y_e&=&\sqrt{P_{s}}\, g_{0}x+\sum_{i=1}^{N} \mathbf{g}_i^T \mathbf{n}_i +n_e.
\end{eqnarray}
The secrecy rate is given by
\al{
R_{2}
&= \log \left( 1+\frac{\gamma_0|h_0|^2}{\sum_{i=1}^N  \gamma_i \qh_i^T{\bs \Sigma}_i \qh_i^\ast +  1}\right)\nnb\\
&\qquad -\log \left( 1 + \frac{\gamma_0|g_0|^2}{\sum_{i=1}^N  \gamma_i \mb{g}_i^T{\bs \Sigma}_i \mb{g}_i^\ast +  1}\right)
\label{Eq:R2}
}
where $\gamma_0$ and $\gamma_i$'s are defined in (\ref{gamma_0}) and (\ref{gamma_i}).
If ${\bs \Sigma}_i = \beta_i\qh_i^\perp (\qh_i^\perp)^\dagger$, it is exactly the orthogonal artificial noise case in (\ref{orthogonal}).

The problem is formulated as
\al{
&\max_{{\bs \Sigma}_i} \ R_{2} \label{Rs2max}\\
&\mathrm{s.t.}\quad {\bs \Sigma}_i \succeq 0, \ \mathrm{and} \ \mathrm{Tr}({\bs \Sigma}_i) \le 1, \ i=1, \cdots, N. \nnb
}
Next we solve the problem of (\ref{Rs2max}).
Using the method in \cite{Jiangyuan1}, let $z = \sum_{i=1}^N  \gamma_i \qh_i^T{\bs \Sigma}_i \qh_i^\ast$.
Note that $\qh_i^T{\bs \Sigma}_i \qh_i^\ast \le \|\qh_i^\ast\|^2\lambda_{\max}({\bs \Sigma}_i) \le \|\qh_i^\ast\|^2\mathrm{Tr}({\bs \Sigma}_i) \le
\|\qh_i^\ast\|^2 = \|\qh_i\|^2$
and equality holds if ${\bs \Sigma}_i = \qh_i^\ast\qh_i^T/\|\qh_i\|^2$.
Thus, the domain of $z$ is
\equ{
0 \le z \le \sum\nolimits_{i=1}^N \gamma_i \|\qh_i\|^2.
}
Obviously, for fixed $z$ (i.e., $\sum_{i=1}^N  \gamma_i \qh_i^T{\bs \Sigma}_i \qh_i^\ast$ is kept fixed at $z$), the term
$\sum_{i=1}^N  \gamma_i \mb{g}_i^T{\bs \Sigma}_i \mb{g}_i^\ast$ should be maximized. With this observation, we let
\al{
F(z) \triangleq \ &\max_{{\bs \Sigma}_i} \ \sum_{i=1}^N  \gamma_i \mb{g}_i^T{\bs \Sigma}_i \mb{g}_i^\ast \label{F}\\
&\mathrm{s.t.}\quad {\bs \Sigma}_i \succeq 0, \ \mathrm{and} \ \mathrm{Tr}({\bs \Sigma}_i) \le 1, \ i=1, \cdots, N;\nnb\\
&\qquad \sum_{i=1}^N  \gamma_i \qh_i^T{\bs \Sigma}_i \qh_i^\ast = z. \nnb
}
The problem of (\ref{F}) is a convex problem which can be effectively solved by CVX \cite{Grant}.
Moreover, we have the following result.

\medskip

\begin{lemma}\label{Lem:Fconcave}
{\em $F(z)$ is a concave function of $z$.}
\end{lemma}
The proof is given in Appendix \ref{proofFconcave}. From the analysis above, the problem of (\ref{Rs2max}) becomes
\al{
&\max_{z}\ R_{2}(z) = \log \Big( 1+\frac{\gamma_0 |h_0|^2}{z +  1} \Big) -\log \Big( 1+\frac{\gamma_0|g_0|^2}{F(z) +  1}\Big), \nnb\\
&\mathrm{s.t.}\quad 0 \le z \le \sum\nolimits_{i=1}^N \gamma_i \|\qh_i\|^2. \label{Rsz-max}
}
The problem of (\ref{Rsz-max}) is equivalent to
\equ{
\max_{z} \left[ G(z) = \frac{1+\frac{\gamma_0 |h_0|^2}{z +  1}}{1+\frac{\gamma_0|g_0|^2}{F(z) +  1}}\right] \ \
\mathrm{s.t.}\ 0 \le z \le \sum_{i=1}^N \gamma_i \|\qh_i\|^2.
}
Since $F(z)$ is a concave function, using the result in \cite[Theorem 3]{Gan}, we have the following result.
\medskip
\begin{theorem}\label{Theo:Gquasiconcave}
{\em $G(z)$ is quasi-concave function of $z$.
}
\end{theorem}

According to Theorem \ref{Theo:Gquasiconcave} and the properties of one variable quasi-concave function \cite[p. 99]{Boyd},
one may determine the global maximizer of $G(z)$ by one dimensional search, e.g., bisection method.

{\em Discussion:} $z=0$ corresponds to the nulling solution with a secrecy rate $R_2(0)=R_1$.
Intuitively, it can be seen from $R_2(z)$ in (\ref{Rsz-max}) that $z$ should not be large,
since the main term $\frac{\gamma_0 |h_0|^2}{z +  1}$ in $R_2(z)$ is $\frac{1}{z+1}$ of $\gamma_0 |h_0|^2$ in $R_1$.
To see this clearly, according to (\ref{Eq:R1Pi}), $R_1$ does not exceed $\log( 1+\gamma_0|h_0|^2)$.
Let $\beta = \frac{R_1}{\log( 1+\gamma_0|h_0|^2)}$.
Solving the following equation for $z$
\equ{
\log\left(1 + \gamma_0|h_0|^2/(z + 1)\right) = \beta \log( 1+\gamma_0|h_0|^2),
}
we get $z = \frac{\gamma_0|h_0|^2}{(1+\gamma_0|h_0|^2)^{\beta} - 1} - 1$.
Thus, 
\equ{
0\le z^\star < \frac{\gamma_0|h_0|^2}{(1+\gamma_0|h_0|^2)^{\beta} - 1} - 1.
}
We can seen that if $\beta$ is near $1$, $z^\star$ will be bounded by a small value compared with $1$.
According to (\ref{Eq:R1Pi}), $\beta$ is near $1$ when $\sum_{i=1}^{N} \gamma_i \mathbf{g}_i^T \mathbf{h}_i^\perp(\mathbf{h}_i^\perp)^\dagger  \mathbf{g}_i^\ast$ is sufficient large compared with $1$. This condition can be satisfied by either more relays (i.e., large $N$) or higher $\gamma_i$.
In numerical simulation section, we give examples which shows that the nulling solution
the nulling solution achieves good approximation to the optimal solution in the average sense.

\section{Simulation and Analysis}\label{sec:Simulations}

In this section numerical simulations are provided to illustrate the nulling scheme and compare it to the general noise scheme.
For illustration purposes, we consider a Gaussian wiretap channel where there are $N=5$ relays, $h_0=0.24 + 0.78i$, $g_0=1.12 - 1.15i$.
For this case, $g_0$ is much stronger than $h_0$, i.e., the eavesdropper channel direct link is much stronger than the legitimate channel direct link;
The artificial noise help to achieve positive secrecy rate.
$\qh_i$'s and $\qg_i$'s are drawn from a Gaussian $\mathcal{CN}(0,\qI)$, for an instance, \\

{\small $\qH =\left(
         \begin{array}{c}
           \qh_1^T \\
           \qh_2^T \\
           \qh_3^T \\
           \qh_4^T \\
           \qh_5^T \\
         \end{array}
       \right)=
$
$\begin{pmatrix}
0.76 - 0.64i & -0.10 - 0.84i\\
-1.077 + 1.15i & -0.96 - 0.18i\\
0.28 + 0.09i & -0.03 - 0.65i\\
0.55 + 0.69i &-0.03 + 0.23i\\
0.39 + 0.01i & -0.82 + 0.27i
\end{pmatrix}$},

{\small $\qG=\left(
       \begin{array}{c}
         \qg_1^T \\
         \qg_2^T \\
         \qg_3^T \\
         \qg_4^T \\
         \qg_5^T \\
       \end{array}
     \right)=
$
$\begin{pmatrix}
0.22 - 0.03i&0.88 + 0.15i\\
-0.165 - 0.29i&0.24 + 0.77i\\
1.10 - 0.47i&0.77 + 0.27i\\
0.33 + 0.79i&0.20 - 0.24i\\
0.88 - 0.05i&0.52 - 0.50i\\
\end{pmatrix}$}.

\begin{figure}[hbtp]
\centering
\includegraphics[width=2.9in]{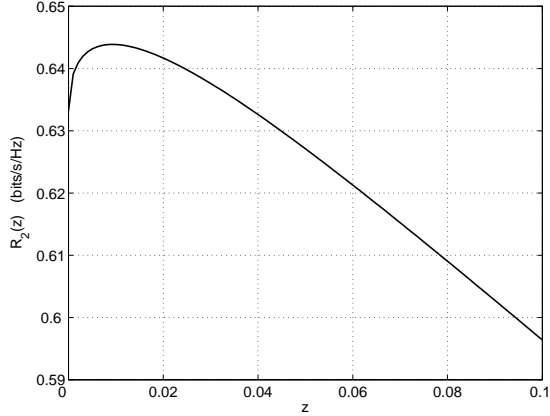}
\caption{$R_2(z)$ vs $z$.}
\label{fig2}
\end{figure}

\begin{figure}[hbtp]
\centering
\includegraphics[width=2.9in]{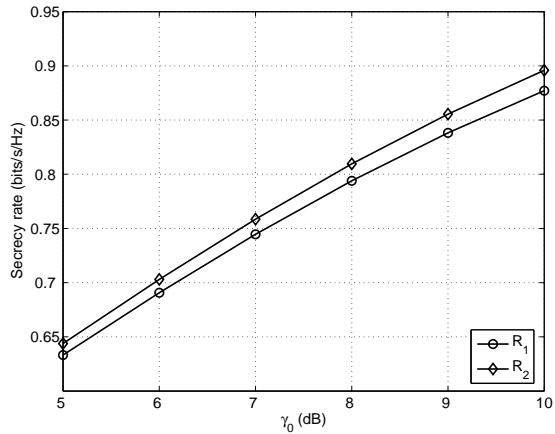}
\caption{$R_1$ and $R_2$ for different values of $\gamma_0$.}
\label{fig3}
\end{figure}

\begin{figure}[hbtp]
\centering
\includegraphics[width=2.9in]{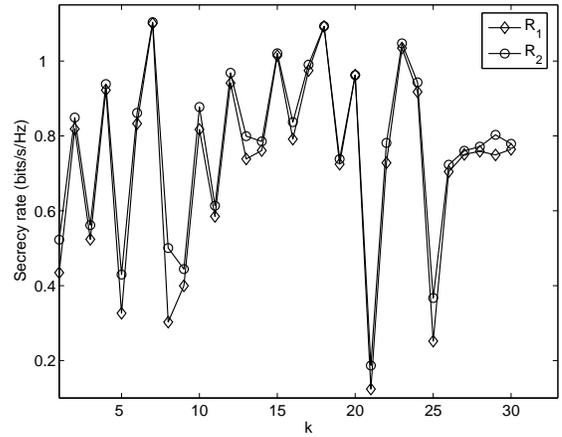}
\caption{$R_1$ and $R_2$ for $30$ randomly generated $\qg_i$'s.}
\label{fig4}
\end{figure}

Let SNRs at source and 5 relays be $(\gamma_0, \gamma_1, \cdots, \gamma_5)=(5, 2, 2, 2, 2, 2)\ \mathrm{dB}$.
The domain of $z$ is $[0, 7.1278]$. The secrecy rate of nulling solution is $R_1=0.6332$.
Fig. \ref{fig2} plots the function $R_2(z)$ over part of the domain of $z$.
The optimal $z$ is $z^\star = 0.0091$ with the objective value $R_2=0.6439$.
It can be seen that $R_1\approx R_2$, i.e., the nulling solution is near optimal.
Then we vary $\gamma_0$ from $5 \, \mathrm{dB}$ to $10 \, \mathrm{dB}$.
Fig. \ref{fig3} plots $R_1$ and $R_2$ for different values of SNR.
Third, we consider the case $\qg_i$'s are drawn from $\mathcal{CN}(0, \qI)$.
Fig. \ref{fig4} plots $R_1$ and $R_2$ for $30$ samples of $\qg_i$'s.
It can be seen from Fig. \ref{fig3} and \ref{fig4} that the nulling solution
achieves good approximation of the optimal solution. For some channel conditions, the nulling solution is near optimal.

\section{Conclusions}
\label{sec:Conclusions}

In this paper, we propose a optimal structured Gaussian noise to do cooperative jamming, which results the maximal secrecy rate. This provides a benchmark for nulling solution, so we compared with local nulling with two antennas per helper and show that for certain SNR range, the optimal noise performs almost the same as the nulling noise. This is particularly encouraging, as  the nulling noise  is very simple to obtain in a distributed manner and without eavesdropper channel knowledge, as opposed to the general noise that requires centralized processing.


\appendices

\section{Proof of Lemma \ref{Lem:Fconcave}}\label{proofFconcave}

We need to prove that, for any feasible $z_1$ and $z_2$
\equ{
F(t z_1 + (1-t)z_2) \ge t F(z_1) + (1-t)F(z_2), \ \forall t\in [0, 1].\label{Convex}
}

Let $\{{\bs \Sigma}_i^{(1)}\}$ and $\{{\bs \Sigma}_i^{(2)}\}$ be the optimal $\{{\bs \Sigma}_i\}$ associated with $z_1$ and $z_2$, respectively.
Consider the problem associated with $t z_1 + (1-t)z_2$, i.e.,
\al{
&\max_{{\bs \Sigma}_i} \ \sum_{i=1}^N  \gamma_i \mb{g}_i^T{\bs \Sigma}_i \mb{g}_i^\ast \label{F-2}\\
&\mathrm{s.t.}\quad {\bs \Sigma}_i \succeq 0, \ \mathrm{and} \ \mathrm{Tr}({\bs \Sigma}_i) \le 1, \ i=1, \cdots, N; \nnb\\
&\qquad \sum_{i=1}^N  \gamma_i \qh_i^T{\bs \Sigma}_i \qh_i^\ast = t z_1 + (1-t)z_2. \nnb
}
It is easy to verify that $\{t{\bs \Sigma}_i^{(1)}+ (1-t){\bs \Sigma}_i^{(2)} \}$ is feasible
with the corresponding objective value $t F(z_1) + (1-t)F(z_2)$.
Thus, (\ref{Convex}) holds and $F(z)$ is concave. This completes the proof.


\begin{thebibliography}{99}
\bibitem{Wyner75} A. D. Wyner, \textquotedblleft The wiretap
channel,\textquotedblright\ \emph{Bell System Technical Journal}, vol. 54,
no. 8, pp. 1355--1387, 1975.

\bibitem{poor} Y. Liang, H. V. Poor and S. Shamai, \emph{Information
Theoretic Security}, Dordrecht, The Netherlands: Now Publishers, 2009.

\bibitem{Dong2} D. Lun, A. P. Petropulu, and H. V. Poor, "Weighted Cross-Layer Cooperative Beamforming for Wireless Networks," Signal Processing, IEEE Transactions on, vol. 57, pp. 3240-3252, 2009.

\bibitem{Jiangyuan1} J. Li, A. P. Petropulu, and S. Weber, ``On cooperative relaying schemes for wireless physical layer security,'' {\em IEEE Trans. Signal Process.}, vol. 59, no. 10, pp. 4985-4997, Oct. 2011.

\bibitem{Gan} G.Zheng, L.C.Choo, and K.K.~Wong, "Optimal cooperative jamming to enhance physical layer security using relays," {\em IEEE Trans. Signal Process.}, vol. 59, no. 3, pp. 1317-1322, Mar. 2011.

\bibitem{Swindlehurst} J. Wang and A. L. Swindlehurst, "Cooperative Jamming
in MIMO Ad-Hoc Networks," {\em Proc. 43rd Asilomar Conference on
Signals, Systems and Computers}, Pacific Grove, CA, Nov. 2009.

\bibitem{Goel1} Negi, R.; Goel, S.; , "Secret communication using artificial noise," {\em Vehicular Technology Conference}, 2005. VTC-2005-Fall. 2005 IEEE 62nd , vol.3, no., pp. 1906- 1910, 25-28 Sept., 2005

\bibitem{Grant} M.Grant and S.Boyd, {\em cvx Users' Guide},
[Online]. Available: http://cvxr.com/, 2009.

\bibitem{Boyd} S.~Boyd and L.~Vandenberghe, {\em Convex Optimization}, Cambridge, UK: Cambridge
Univ. Press, 2004.

\end{thebibliography}
\end{document}